\documentclass[aps,prl,amsbsy,floatfix,twocolumn,sort&compress]{revtex4}

\usepackage{amssymb} 
\usepackage{graphicx} 
\usepackage{bm} 

\newcommand{\gtwid}{\mathrel{\raise.3ex\hbox{$>$\kern-.75em\lower1ex\hbox{$\sim$}}}} 
\newcommand{\ltwid}{\mathrel{\raise.3ex\hbox{$<$\kern-.75em\lower1ex\hbox{$\sim$}}}} 
\newcommand{\etal}{{\it et al.}} 
\newcommand*\Bell{\ensuremath{\boldsymbol\ell}} 

\begin{document}

\title{Pair Structure and the Pairing Interaction in a Bilayer Hubbard model} 
\date{\today } 
\author{T.A.~Maier$^{1}$ and D.~J.~Scalapino$^{2}$\\
{$^1$ Computer Science and Mathematics Division and Center for Nanophase Materials Sciences, Oak Ridge National Laboratory, Oak Ridge, Tennessee 37831-6494}\\
{$^2$ Department of Physics, University of California, Santa Barbara, CA 93106-9530, USA}}

\begin{abstract}
The bilayer Hubbard model with an intra-layer hopping $t$ and an inter-layer hopping $t_\perp$ provides an interesting testing ground for several aspects of what has been called unconventional superconductivity. One can study the type of pair structures which arise when there are multiple Fermi surfaces. One can also examine the pairing for a system in which the structure of the spin-fluctuation spectral weight can be changed. Using a dynamic cluster quantum Monte Carlo approximation, we find that near half-filling, if the splitting between the bonding and anti-bonding bands $t_\perp/t$ is small, the gap has $B_{1g}$ ($d_{x^2-y^2}$-wave) symmetry but when the splitting becomes larger, $A_{1g}$ ($s^\pm$-wave) pairing is favored. We also find that in the $s^\pm$ pairing region, the pairing is driven by inter-layer spin fluctuations and that $T_c$ is enhanced. 
\end{abstract}

\maketitle

The bilayer Hubbard model provides a model system for which one can study the relationship between the structure of the Fermi surface, the spin-fluctuation spectrum and superconductivity. By varying the relative strength of the inter-layer one electron hopping $t_\perp$ to the near-neighbor intra-layer hopping $t$, one can alter the size and shape of the bonding and antibonding Fermi surfaces, change the momentum and frequency structure of the spin-fluctuation spectral weight, and move away from the Mott region. Previous Monte Carlo calculations \cite{ref:3,ref:4,ref:Hanke,ref:5,ref:6,ref:7} for a doped bilayer found evidence for an attractive pairing interaction in both the $d_{x^2-y^2}$ ($\cos k_x-\cos k_y$) and $\cos k_z$ channels. The latter was called a $d_z$ channel and corresponded to a pairfield function which had different signs on the bonding and antibonding Fermi surfaces. This is an $A_{1g}$ gap and here we will refer to it as an $s^\pm$ gap. The sign problem associated with the previous determinantal quantum Monte Carlo calculations prevented one from numerically exploring the low temperature properties of the doped bilayer. More recently, a functional renormalization group study \cite{ref:Lee} found that as $t_\perp/t$ increased, $s^\pm$ pairing was favored over $d_{x^2-y^2}$ pairing. Here using a dynamic cluster approximation\cite{ref:8} (DCA) we will examine what the bilayer Hubbard model tells us about the pairing mechanism and the search for higher $T_c$ materials.

\begin{figure}
	[htbp] 
	\includegraphics[width=1.8in]{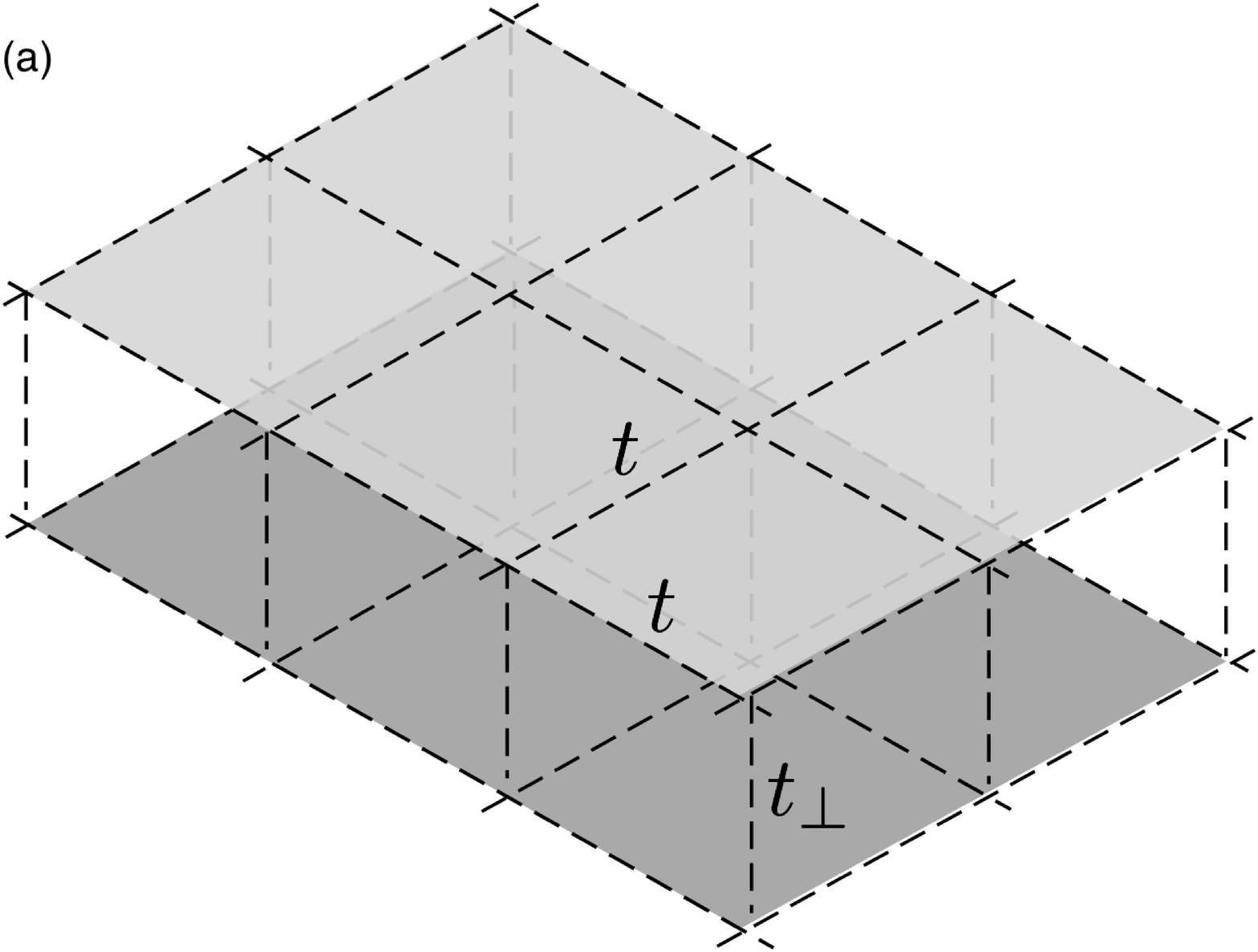}\includegraphics[height=1.8in]{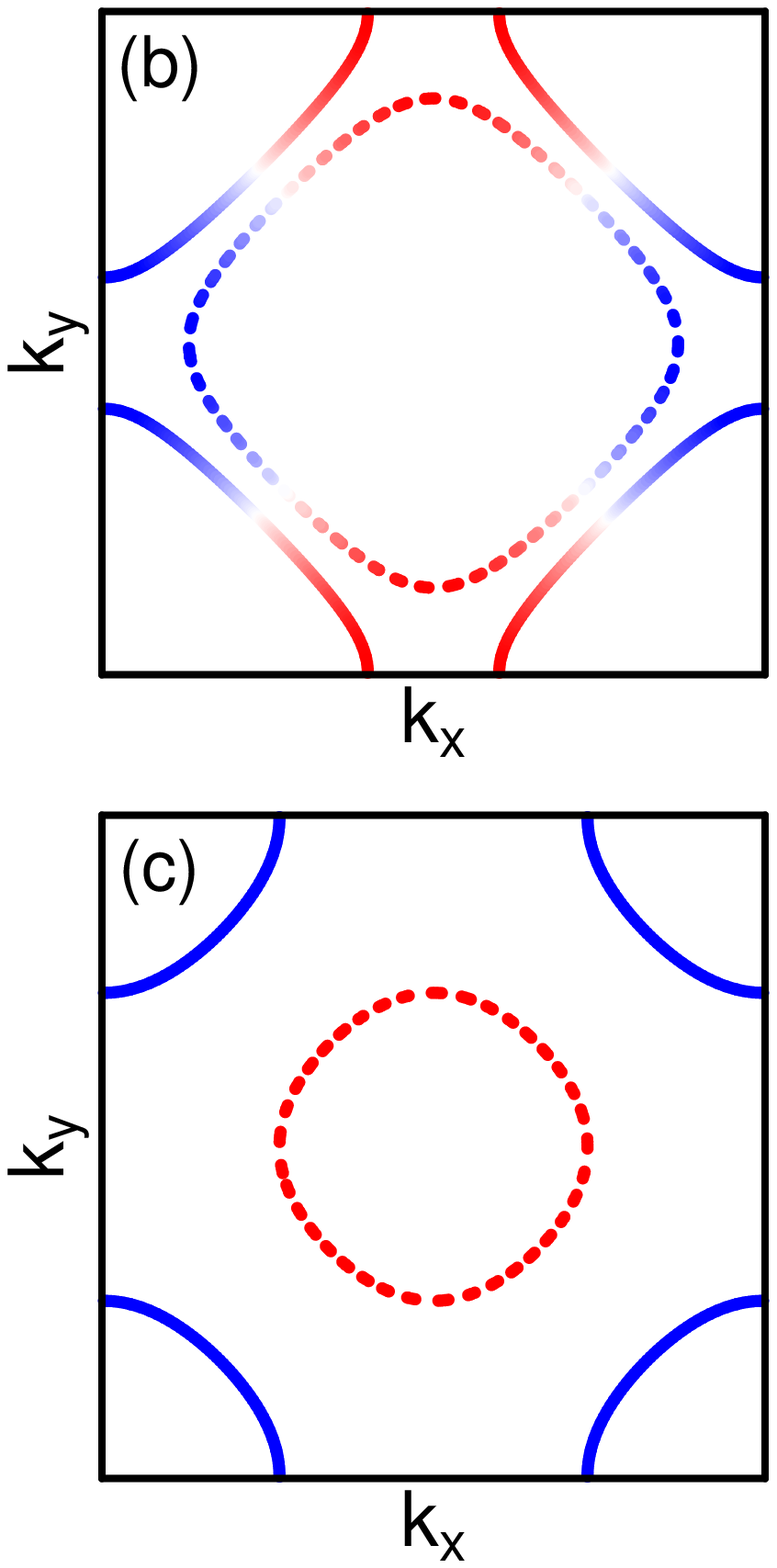} \caption{a) The bilayer Hubbard lattice with near neighbor intra-layer $t$ and inter-layer $t_\perp$ hopping parameters and an onsite $U$ Coulomb interaction. The bonding ($k_z=0$) and antibonding ($k_z=\pi$) Fermi surfaces for $t_\perp=0.5$ (b) and 2.0 (c) at a filling $\langle n\rangle=0.95$. A $d_{x^2-y^2}$ gap structure is illustrated for the $t_\perp/t=0.5$ Fermi surface and an $s^\pm$ gap structure is shown for $t_\perp/t=2.0$. \label{fig:1}} 
\end{figure}

The Hamiltonian for the bilayer Hubbard model shown in Fig.~\ref{fig:1} can be written as 
\begin{eqnarray}
	H&=&-t\sum_{\langle ij\rangle m\sigma}(c^+_{jm\sigma}c_{im\sigma}+{\rm h.c.})- t_\perp\sum_{i\sigma}(c^+_{i1\sigma}c_{i2\sigma}+{\rm h.c.})\nonumber\\
	&-& \mu\sum_{im\sigma}n_{im\sigma}+U\sum_{im}n_{im\uparrow}n_{im\downarrow} \label{eq:1} 
\end{eqnarray}
Here $t$ and $t_\perp$ are intra- and inter-layer near neighbor hopping parameters, $\mu$ is the chemical potential and $U$ is an onsite Coulomb interaction. The indices $i$ and $j$ run over the sites in both the $m=2$ (upper) and $m=1$ (lower) layers and $\langle ij\rangle$ implies that only a single $\langle ij\rangle$ near neighbor hopping is included in the sum. In the following we will measure energies in terms of $t$ and set $U=6$. The bonding $(k_z=0)$ and anti-bonding $(k_z=\pi)$ bands are given by 
\begin{equation}
	\varepsilon(k)=-2t(\cos k_x+\cos k_y)\pm t_\perp\cos k_z \label{eq:2} 
\end{equation}
The $k_z=0$ bonding and $k_z=\pi$ anti-bonding Fermi surfaces (FS) of the non-interacting system are shown in Fig.~\ref{fig:1} for a filling $\langle n\rangle=0.95$ and two values of $t_\perp/t$. There is an anti-bonding ($k_z=\pi$) electron-like FS that forms around the ($k_x=0$, $k_y=0$) $\Gamma$ point of the 2D Brillouin zone and a bonding ($k_z=0$) hole-like FS around the ($\pi,\pi$) point. For $\langle n \rangle=0.95$, the topology of the non-interacting FS changes from two electron-like FS's to one electron and one hole-like FS when $t_\perp/t$ exceeds a critical value of order 0.1. Here we will focus on the behavior of the bilayer system for $t_\perp/t$ greater than this value, although in some plots we will give the $t_\perp=0$ result. As $t_\perp/t$ increases further, the FS's shown in Fig.~1 shrink and for $\langle n\rangle=1$ the non-interacting system becomes a band insulator when $t_\perp > 4t$. Weak coupling fluctuation-exchange (FLEX) \cite{ref:3} as well as phenomenological \cite{ref:9} spin-fluctuation calculations find superconductivity for the doped system with a $d_{x^2-y^2}$-like gap for $t_\perp/t=0.5$ and an $s^\pm$-like gap for $t_\perp/t=2.0$, as schematically illustrated in Fig.~\ref{fig:1}. This two-Fermi-surface model can be seen as an analog of the multi-Fermi-surface Fe-pnictide models. 

In the following, we will use the DCA to study the Hubbard bilayer with $U=6$ and $\langle n\rangle=0.95$. The DCA \cite{ref:8,ref:Jarrell2001} maps the bulk lattice problem onto an effective periodic cluster embedded in a self-consistent dynamic mean-field that is designed to represent the remaining degrees of freedom. The DCA calculations were carried out on an $N=2\times(L\times L)$ cluster with $L=4$ and the effective cluster problem was solved using a Hirsch-Fye quantum Monte Carlo algorithm \cite{ref:Jarrell2001}. 

\begin{figure}
	[htbp] 
	\includegraphics[width=7.5cm]{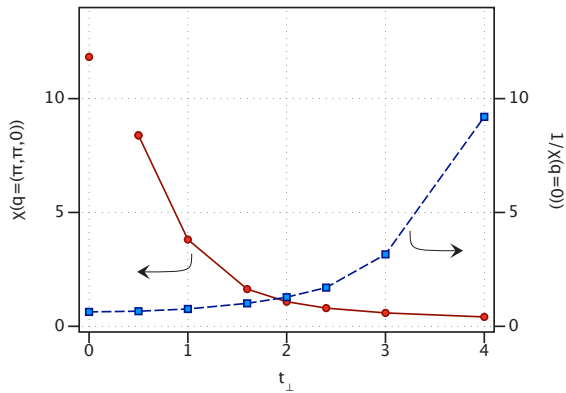} \caption{The spin susceptibility $\chi(\pi,\pi,0)$ and $\chi^{-1}(0)$ for $\langle n\rangle=0.95$, $U=6$ and $T=0.4$ are plotted versus $t_\perp$. As $t_\perp$ exceeds 2, the planar AF response is suppressed and singlet correlations between sites on opposite sides of the bilayer become dominant. \label{fig:2}} 
\end{figure}

The magnetic susceptibility is given by 
\begin{equation}
	\chi({\bf q})=\int^\beta_0d\tau\langle m^-_{\bf q}(\tau)m^\dagger_{\bf q}(0)\rangle \label{eq:3} 
\end{equation}
Here 
\begin{equation}
	m^\dagger_{\bf q}=\frac{1}{\sqrt{N}}\sum_{\bf k}c^+_{{\bf k}+{\bf q}\uparrow}c_{{\bf k}\downarrow} \label{eq:4} 
\end{equation}
and 
\begin{equation}
	c_{{\bf k}\downarrow}=\frac{1}{\sqrt{N}}\sum_{({\Bell},m)}e^{({\bf k}\cdot{\Bell}+k_zm)}c_{{\Bell}m\downarrow} \label{eq:5} 
\end{equation}
Fig.~\ref{fig:2} shows plots of $\chi(\pi,\pi,0)$ and $\chi^{-1}(0)$ versus $t_\perp$ at a temperature $T=0.4$. 
For the undoped system at $T=0$, the AF order vanishes for $t_\perp\gtwid2.0$ and the ground state is a disordered valence bond (VB) phase with a spin gap \cite{ref:7}. Here, for the doped system, one sees that for $t_\perp\sim2$ the near neighbor in plane ($\pi,\pi,0$) response is suppressed and as $t_\perp$ increases further, the low energy interlayer spin-fluctuations become gapped as the interlayer valence-bond singlets form. This crossover is also clearly seen in the behavior of the inverse spin susceptibility $\chi^{-1}(q=0)$. 

\begin{figure}
	[htbp] 
	\includegraphics[width=1.7in]{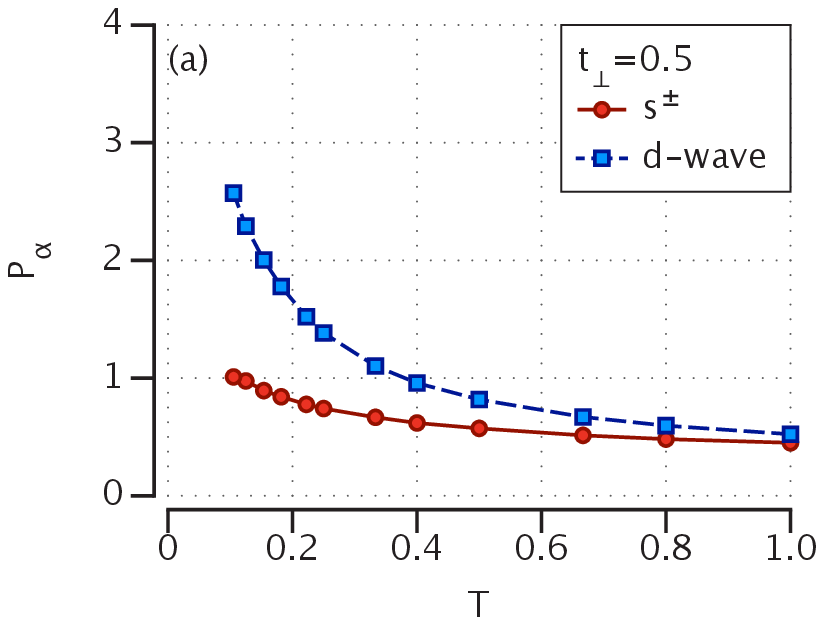}\includegraphics[width=1.7in]{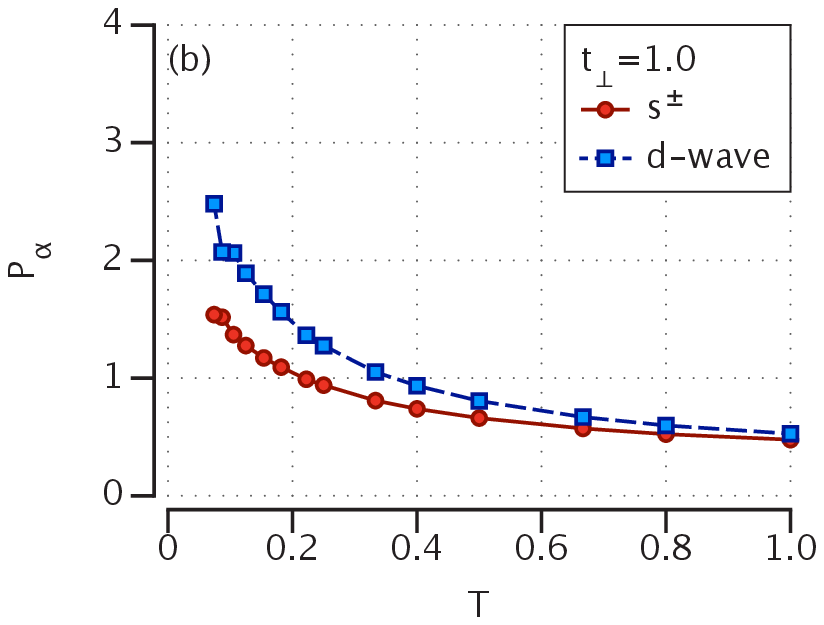}\\
	\includegraphics[width=1.7in]{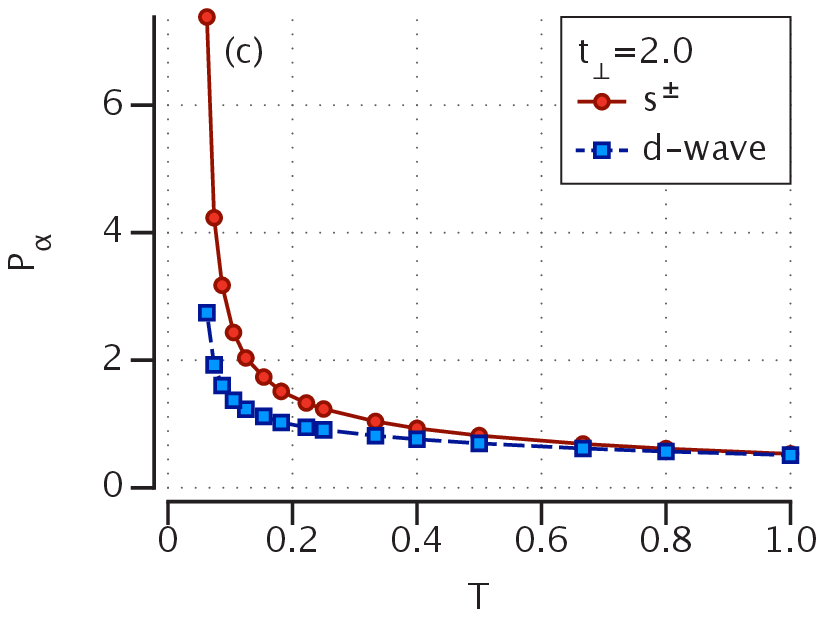}\includegraphics[width=1.7in]{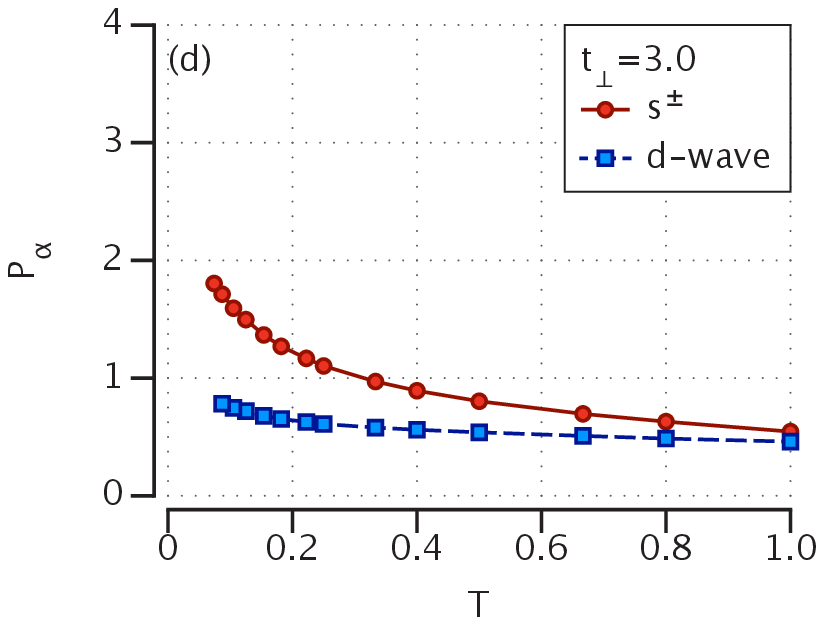} \caption{a) The $s^\pm$ and $d$-wave pairfield susceptibilities $P_\alpha$ versus temperature for different values of the inter-layer hopping $t_\perp$. As $t_\perp$ increases, the leading pairfield susceptibility changes from $d$-wave to $s^\pm$ and at larger values of $t_\perp$ the superconducting pairfield susceptibility is suppressed as inter-layer valence bonds form. \label{fig:3}} 
\end{figure}

As noted, weak coupling calculations have found both $d_{x^2-y^2}$ ($B_{1g}$) and $s^\pm$ ($A_{1g}$) pairing correlations for the Hubbard bilayer model. Here, in order to probe the superconducting response we have used the DCA to calculate the pairfield susceptibilities. 
\begin{equation}
	P_\alpha(T)=\int^\beta_0d\tau\langle\Delta_\alpha(\tau)\Delta^\dagger_\alpha(0)\rangle \label{eq:6} 
\end{equation}
associated with each of these symmetries. For the $d_{x^2-y^2}$-wave we have taken 
\begin{equation}
	\Delta^\dagger_{x^2-y^2}=\frac{1}{\sqrt N}\sum_k(\cos k_x-\cos k_y)c^\dagger_{k\uparrow}c^\dagger_{-k\downarrow} \label{eq:7} 
\end{equation}
and for the $s^\pm$ case \cite{ref:ft}
\begin{equation}
	\Delta^\dagger_{s^\pm}=\frac{1}{\sqrt N}\sum_k\cos k_zc^\dagger_{k\uparrow}c^\dagger_{-k\downarrow} \label{eq:8} 
\end{equation}
Here, $k=(k_x,k_y,k_z)$ with $k_z=0$ and $\pi$. In Fig.~\ref{fig:3}, results for these pairfield susceptibilities are shown for different values of $t_\perp$ versus temperature. 
For $t_\perp=0.5$, one sees in Fig.~\ref{fig:3}a that the $d_{x^2-y^2}$-wave susceptibility is rising the most rapidly at low temperatures followed by the $s^\pm$-wave response. In Fig.~\ref{fig:3}b, for $t_\perp=1.0$ the $d$-wave response is slightly larger than the $s^\pm$ case but they are closer, while for $t_\perp=2.0$ the $s^\pm$ is the leading pair susceptibility. This result is similar to that found in Ref.~\cite{ref:Lee}. The final panel, Fig.~\ref{fig:3}d, shows the pairfield susceptibilities for $t_\perp=3$. For this value of $t_\perp$, the $T=0$ undoped system is well into the spin-gapped VB phase and one sees that the pairfield response of the doped system is significantly weakened.

\begin{figure}
	[htbp] 
	\includegraphics[width=1.7in]{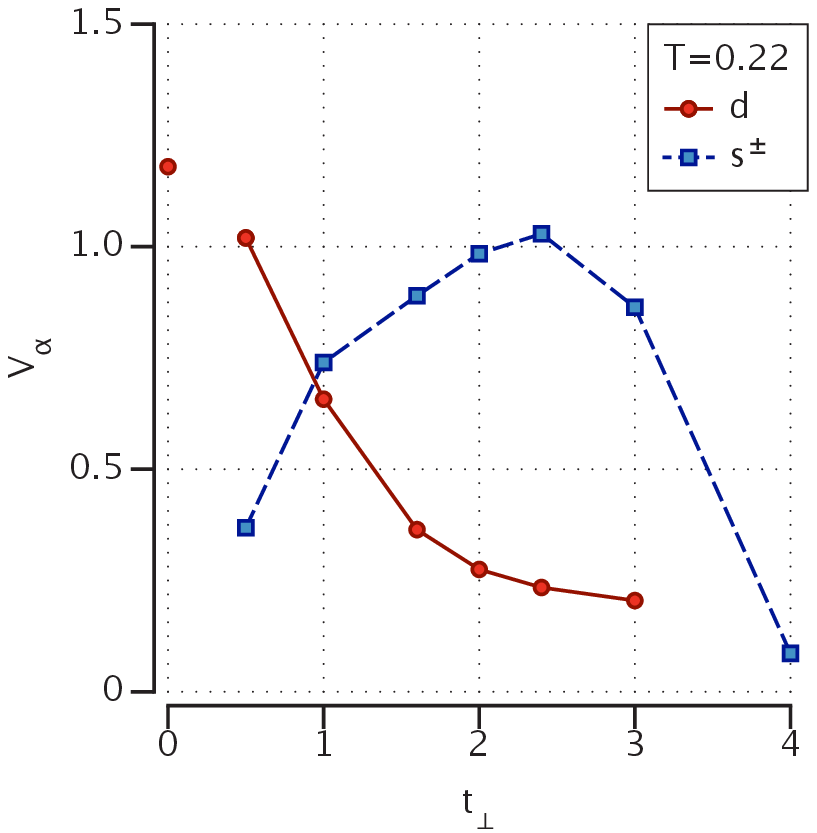}\includegraphics[width=1.7in]{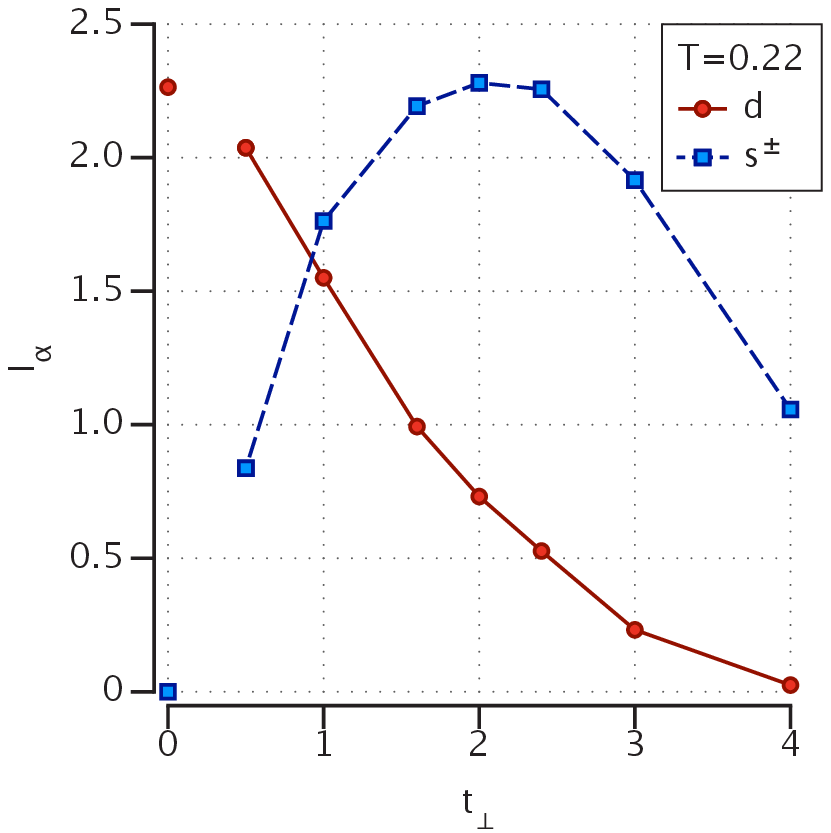} \caption{(a) The strength of the pairing interaction $V_\alpha$ for $\alpha=d$ and $s^\pm$ calculated from Eq.~(\ref{eq:9}) versus $t_\perp$. The integrated spectral weights $I_\nu$, Eq.~(\ref{eq:10}), for the intra- ($\nu=x$) and inter- ($\nu=z$) layer near-neighbor spin-fluctuations versus $t_\perp$. \label{fig:4}} 
\end{figure}

The pairing interaction is given by the irreducible particle-particle scattering vertex $\Gamma(K,K')$ which describes the scattering of a singlet pair from the state $(K,-K)$ to $(K',-K')$ with $K=({\bf K},i\omega_n)$. Given this vertex and the dressed single-particle Green's functions, one can solve the Bethe-Salpeter equation to find the pairing eigenvalues $\lambda_\alpha$ and eigenfunctions $\varphi_\alpha(K)$ \cite{ref:Maier}. For the bilayer Hubbard model, the leading pairing eigenvalue at small values of $t_\perp$ corresponds to the $B_{1g}$ ($d_{x^2-y^2}$) state and, as $t_\perp$ exceeds 1, switches to the $A_{1g}$ ($s^\pm$) state. A measure of the pairing strength $V_\alpha$ for a given pairing state is

\begin{equation}
	V_{\alpha}=\frac{1/N\sum_{{\bf K},{\bf K}'}\varphi_{\alpha}({\bf K},\pi T)\Gamma({\bf K},\pi T,{\bf K}',\pi T) \varphi_{\alpha}({\bf K}',\pi T)}{\sum_{\bf K}\varphi^2_{\alpha}({\bf K},\pi T)} \label{eq:9} 
\end{equation}

A plot of $V_\alpha$ versus $t_\perp$ is shown on the left hand side of Fig.4. On the right hand side of this figure we have plotted the integrated spin-fluctuation spectral weights for the intra- and inter-layer near-neighbor spin fluctuations
\begin{eqnarray} \label{eq:10}
	I_\nu&=&\frac{1}{N}\sum_{\bf K}\int\frac{d\omega}{\pi}\frac{\chi^{\prime\prime}_c({\bf K},\omega)}{\omega}\cos {\bf K}_\nu\nonumber\\
	& =& \frac{1}{N}\sum_{\bf K}\chi^{\prime}_c({\bf K},0) \cos {\bf K}_\nu
\end{eqnarray}
Here ${\bf K}_\nu={\bf K}_x$ for the intra-layer fluctuations and ${\bf K}_z$ for the inter-layer fluctuations, and $\chi_c({\bf K},\omega)$ is the spin-susceptibility calculated on the $2\times(4\times4)$ cluster. As $t_\perp$ increases, the dominant spin-fluctuations change from intra-layer to inter-layer. The inter-layer spin fluctuations give rise to the scattering of pairs between the bonding and anti-bonding Fermi surfaces and lead to $s^\pm$ pairing. We believe that the similarity of the $t_\perp$ dependence of $I_\nu$ and the pairing strengths $V_\alpha$ provides evidence linking the structure of the spin fluctuation spectral weight to the pairing mechanism and the resulting gap structure in this model. This spin-fluctuation picture differs from an alternate strong coupling scenario in which $U$ is assumed larger than the bandwidth and superconductivity is viewed as arising from the melting of a valence bond phase with doping \cite{ref:10}. We believe that the spin-fluctuation picture provides a more appropriate framework for $U$ of order the bandwidth, which corresponds to the parameter region in which the pairing is strongest.

Although we cannot carry out the finite size scaling analysis one would need to determine $T_c$, we have estimated $T_c$ values from the temperature at which the DCA results for $P_\alpha(T)$ diverge. We find that the maximum $T_c$ for the $s^\pm$ phase is significantly larger than that for the $d$-wave phase even though, as seen in Fig.~4, the maximum value of $V_d$ exceeds that of $V_{s^\pm}$. This is because, at $\langle n \rangle=0.95$, the $t_\perp\rightarrow 0$ system is near the Mott state and the quasi-particle weight is strongly suppressed. At larger values of $t_\perp$, where $V_{s^\pm}$ peaks, the bilayer system with an intermediate $U=6$ has moved into a semi-metallic phase which has only a moderate renormalization of the quasi-particles. It is interesting to note that the maximum $T_c$ for the $s^\pm$ phase with $\langle n\rangle=0.95$ is more than a factor of two larger than the $T_c$ obtained from a corresponding DCA estimate of the single-layer case at optimal doping. In this case, the $T_c$ of the single-layer system has been increased by doping away from half-filling so as to suppress the quasi-particle Mott renormalization. However, this reduces the strength of the pairing interaction. In the bilayer case, one can move away from the Mott region by increasing $t_\perp$ and at the same time still achieve a large interaction strength as shown in Fig.4. Thus we conclude that multi Fermi surface materials can offer a possible pathway to higher $T_c$ superconductivity.

To conclude, in this paper we have reported DCA results for the pair structure and pairing interaction for a bilayer Hubbard model. By increasing the relative size of the inter-layer hopping $t_\perp$ to the intra-layer hopping $t$ at a fixed doping, we have shown that one can tune the system from a $d_{x^2-y^2}$ (B$_{1g}$) superconductor to an $s^\pm$ (A$_{1g}$) superconductor. We then examined the strength of the pairing interactions in these channels as $t_\perp/t$ was varied and showed that they were correlated with the relative strengths of the near-neighbor intra-layer to inter-layer spin fluctuations respectively. When the inter-layer spin fluctuations become dominant, an $s^\pm$ superconducting phase is favored. The ability to change the nature of the superconducting state by varying a one-electron parameter in the Hamiltonian represents a useful approach for studying the relationship between the pair structure and the underlying pairing interaction as well as identifying the basic correlations which are responsible for pairing.

We would like to acknowledge the Center for Nanophase Materials Sciences, which is sponsored at Oak Ridge National Laboratory by the Office of Basic Energy Sciences, U.S. Department of Energy. This research was enabled by computational resources of the Center for Computational Sciences at Oak Ridge National Laboratory. D.J.S. would like to thank the Stanford Institute of Theoretical Physics for their hospitality.

\end{document}